\documentclass[11pt, notitlepage]{article}
\pdfoutput=1

\usepackage{url} 
\usepackage{float}
\usepackage{subfig}
\usepackage{graphicx}
\usepackage[T1]{fontenc}
\usepackage[utf8]{inputenc}
\usepackage{authblk}
\usepackage{amsmath,bm}
\usepackage{amssymb}
\usepackage{amsmath}
\usepackage{enumitem}
\usepackage{mathtools}
\usepackage{authblk}
\usepackage[left=2cm,right=2cm,top=2cm,bottom=2cm]{geometry}

\renewcommand\footnotemark{}
\begin{document}

\title{\bf{A Falsifiable Alternative to General Relativity}}
\author{Daniel Coumbe}
\author[2]{Aria Rahmaty}
\thanks{E-mail: DC@nrgym.dk}
\affil[1]{\small{\emph{The Niels Bohr Institute, Copenhagen University}\\ \emph{Blegdamsvej 17, DK-2100 Copenhagen Ø, Denmark}}}
\affil[2]{\small{\emph{Nørre Gymnasium, Mørkhøjvej 78, 2700 Brønshøj, Denmark}}}
\date{}
\maketitle


\begin{abstract}

  Asymptotically Weyl-invariant gravity (AWIG) is further developed within the Palatini formalism as a power-counting renormalizable alternative to general relativity (GR). An expression for the dimensionless exponent $n\left(\mathcal{R}\right)$ is derived based on dynamical dimensional reduction. We show that this version of AWIG naturally resolves several theoretical issues normally associated with the Palatini formalism. A falsifiable prediction regarding the frequency of gravitational waves from binary black hole mergers is made. A preliminary analysis of gravitational wave GW150914 yields a maximum tension of $0.9 \sigma$ with GR and marginally favours AWIG. A similar analysis of gravitational wave GW151226 yields a maximum tension of $2.7 \sigma$ with GR and favours AWIG more significantly.  

\vspace{0.25cm}

\end{abstract}


\begin{section}{Introduction}

  Einstein's general theory of relativity has reigned supreme for over a century as our best description of gravity. However, we now understand that it must be modified or replaced entirely.

 The need to modify general relativity arises from purely theoretical considerations. A significant problem with the Einstein-Hilbert action is its incompatibility with quantum field theory at high energies. While it is true that general relativity has been successfully formulated as an effective quantum field theory at low energy scales, at higher energies new divergences appear at each order in the perturbative expansion, ultimately leading to a complete loss of predictivity. Gravity is thus said to be perturbatively non-renormalizable, as demonstrated by explicit calculation at the one-loop level including matter content~\cite{'tHooft:1974bx}, and at the two-loop level for pure gravity~\cite{Goroff:1985th}. Moreover, general relativity predicts its own breakdown at curvature singularities within black holes or at the beginning of the universe, where scalar measures of curvature grow without bound. Recently, this theoretical problem became very real due to the first direct image of a black hole~\cite{EventHorizonTelescope:2022wkp}, as well as the first indirect observation of merging black holes via gravitational waves~\cite{LIGOScientific:2016aoc}.     

 Further motivation for modifying general relativity comes from the experimental sector. Astrophysical and cosmological observations indicate that retaining the standard gravitational action of general relativity requires the existence of undetected non-baryonic matter (dark matter) and an unknown form of energy (dark energy) comprising approximately $27\%$ and $68\%$ of the universe's energy budget, respectively~\cite{Planck:2018vyg}. Moreover, various cosmological puzzles such as the horizon, flatness, and monopole problems indicate an early period of rapid cosmic expansion. One model that attempts to fit all of this data is $\Lambda$CDM (Lambda cold dark matter) supplemented by an inflationary scenario~\cite{Sotiriou:2008rp}. In addition to being plagued by the cosmological constant and coincidence problems, the $\Lambda$CDM model fails to explain the origin of the inflaton field or the nature of dark matter. Since $\Lambda$CDM is more of an ad-hoc empirical fit to astrophysical data, rather than a theoretically well-motivated theory, it has led many to consider modifying our description of gravity~\cite{Hess:2020ssc}.  

 Perhaps the simplest way of modifying Einstein's general theory of relativity is $f(R)$ gravity~\cite{Capozziello_2011}. In $f(R)$ gravity, the specific Lagrange density $R$ of the Einstein-Hilbert action is replaced by a general function $f(R)$, where $R$ is the scalar curvature. Hence, the action becomes

\begin{equation}\label{02}
S=\frac{1}{\kappa} \int R \sqrt{-g} d^{4}x\hspace{0.5cm} \to\hspace{0.5cm} S_{f}=\frac{1}{\kappa} \int f(R) \sqrt{-g} d^{4}x, 
\end{equation} 

\noindent where $\kappa=16\pi G$, $G$ is Newton's constant and $g$ is the determinant of the metric tensor. A key advantage of $f(R)$ theories is their evasion of Ostragadsky's powerful no-go theorem, as they uniquely violate Ostragadsky’s assumption of non-degeneracy~\cite{Woodard:2015zca,Woodard:2006nt}. There are three types of $f(R)$ gravity: metric, metric-affine, and the Palatini formalism~\cite{Sotiriou:2008rp}.

Metric $f(R)$ gravity assumes that the affine connection uniquely depends on the metric via the Levi-Civita connection, as in standard general relativity. \emph{A priori} there is no logical reason for this assumption. Metric $f(R)$ models also typically suffer from several problems. For example, certain metric $f(R)$ models are known to conflict with solar system tests~\cite{Chiba:2003ir}, give an incorrect Newtonian limit~\cite{Sotiriou:2005hu}, contradict observed cosmological dynamics~\cite{Amendola:2006kh,Amendola:2006we}, be unable to satisfy big bang nucleosynthesis constraints~\cite{Brookfield:2006mq} and contain fatal Ricci scalar instabilities~\cite{Dolgov:2003px}. Meanwhile, models defined within the metric-affine formalism of $f(R)$ gravity do not even constitute metric theories, meaning the fundamental symmetry of diffeomorphism invariance is likely broken~\cite{Sotiriou:2008rp}.

This leaves the Palatini formalism, which, despite its name, was invented by Einstein in 1925. A particular strength of the Palatini formalism is that the Levi-Civita connection is not assumed \emph{a priori}, as it is in the metric formalism, but arises as a natural consequence of the dynamics of the action. Palatini $f(R)$ gravity is also immune to many of the problems inherent in the other two formulations. For example, Palatini gravity has no Ricci scalar instability~\cite{Meng:2003sxc} and is strictly a metric theory and thus diffeomorphism invariant~\cite{Sotiriou:2005xe}. Remarkably, a Palatini action that is linear in the scalar curvature also turns out to be identical to standard general relativity~\cite{Sotiriou:2008rp}. Moreover, since this equivalence does not hold for actions non-linear in the scalar curvature it opens up the exciting possibility for new cosmological dynamics at higher curvature scales~\cite{BeltranJimenez:2017vop,Sotiriou:2008rp}. Despite these advantages, the Palatini formalism still has several outstanding issues. In this work, we show that our particular formulation of Palatini gravity can solve many, if not all, of these problems. 

In Ref.~\cite{Coumbe:2019fht}, the theory of asymptotically Weyl-invariant gravity (AWIG) was proposed within the Palatini formalism (see Refs.~\cite{Coumbe:2015zqa,Coumbe:2015aev,Coumbe:2018myj} for the motivation behind this proposal). This theory is defined by the action

 \begin{equation}\label{03}
S_{p}=\frac{1}{\kappa} \int \mathcal{R}^{n(\mathcal{R})}\sqrt{-g}d^{4}x,\qquad n(\mathcal{R})\to
\begin{cases}
      1,&\ \text{for}\ \mathcal{R}\to 0 \\
      2,&\ \text{for}\ \mathcal{R}\to \infty,
    \end{cases}
\end{equation}

 \noindent where $\mathcal{R}$ is the Ricci scalar in the Palatini formalism given by $\mathcal{R}=g^{\mu\nu}\left(\partial_{\rho}\Gamma^{\rho}_{\nu\mu} - \partial_{\nu}\Gamma^{\rho}_{\rho\mu} + \Gamma^{\rho}_{\rho\lambda}\Gamma^{\lambda}_{\nu\mu} - \Gamma^{\rho}_{\nu\lambda}\Gamma^{\lambda}_{\rho\mu}\right)$ with the metric $g_{\mu\nu}$ and connection $\Gamma^{\sigma}_{\mu\nu}$ treated as independent variables~\cite{Edery:2019txq}. The exponent $n(\mathcal{R})$ is a dimensionless function of $\mathcal{R}$~\cite{Coumbe:2019fht}. In the low-curvature limit (IR) AWIG recovers $f \left(\mathcal{R}\right)=\mathcal{R}$, which is identical to general relativity~\cite{Sotiriou:2008rp}. In the high curvature limit (UV) AWIG asymptotically approaches $f \left(\mathcal{R}\right)=\mathcal{R}^{2}$, which is a power-counting renormalizable and locally scale-invariant theory~\cite{Coumbe:2019fht,Coumbe:2021qid}.\interfootnotelinepenalty=10000 \footnote{\scriptsize Presently, we only claim that AWIG is superficially renormalizable as shown by simple power-counting arguments. Namely, since gravity as a perturbative quantum field theory in $d$-dimensional spacetime dictates that the momentum $p$ scales with loop-order $L$ via $\int p^{A-[G]L} dp$~\cite{Weinberg79}, where $[G]$ is the canonical mass dimension of Newton's coupling and $A$ is a process-dependent constant that is independent of $L$, then because $[G]\to 0$ in the UV limit of AWIG the momenta $p$ no longer scales with $L$, making it at least power-counting renormalizable.} AWIG is likely to be unitary because states of negative norm (ghosts) that cause unitarity violations do not appear in $f(R)$ theories~\cite{Sotiriou:2008rp,Stelle:1977ry}. 
 
 The fundamental symmetry behind AWIG is that of local scale invariance (LSI). The need for LSI can be motivated by considering the foundational question of what it means to make a measurement. For example, all length measurements are local comparisons. To measure the length of a rod requires bringing it together with some standard unit of length, say a metre stick, at the same point in space and time. In this way, the local comparison yields a dimensionless ratio. Repeating this comparison at a different spacetime point must yield the same ratio, even if the metric at this new point were rescaled by an arbitrary factor $\Omega^{2}(x)$. As a guiding principle, we therefore propose that the laws of nature must be formulated in such a way as to be invariant under local rescalings of the metric tensor $g_{\mu\nu} \rightarrow \Omega^{2}(x) g_{\mu\nu}$, \emph{$\grave{a}$ la} Weyl invariance. Since scale-invariant theories of gravity are gauge theories~\cite{Wesson,Lasenby:2015dba}, unification with the other three fundamental forces, which have all been successfully formulated as local gauge theories, may become tractable.

\end{section}

  
\begin{section}{Completing the model}\label{sec2}

  In this section we aim to complete the definition of AWIG by deriving the functional form of the as yet undefined exponent $n\left(\mathcal{R}\right)$ in Eq.~(\ref{03}).

In $N-$dimensions, all locally scale-invariant $f(\mathcal{R})$ actions in the Palatini formalism take the form~\cite{Dabrowski:2008kx,Coumbe:2019fht}

\begin{equation}\label{a1}
S=\frac{1}{\kappa} \int \mathcal{R}^{N/2} \sqrt{-g} d^{N}x.
\end{equation} 

\noindent Consider a generalised version of Eq.~(\ref{a1}) of the form 

\begin{equation}\label{a2}
S=\frac{1}{\kappa} \int \mathcal{R}^{a/2} \sqrt{-g} d^{b}x,
\end{equation}  
  
\noindent where the two parameters $a$ and $b$ are not necessarily equal. Expressing the action in this way quantifies the degree to which local scale invariance is broken via departures from the Weyl symmetric case $a=b$. A dimensionless action requires $[\kappa]=[\mathcal{R}^{a/2}]+[\sqrt{-g}]$, and since the canonical length dimension of $\mathcal{R}^{a/2}$ is $-a$ and for $\sqrt{-g}$ it is $b$, this implies $[\kappa]=b-a$.\interfootnotelinepenalty=10000 \footnote{\scriptsize Here we have chosen the convention in which the metric tensor is assigned a dimensionality rather than the coordinates, which one is free to do since it is a matter of convention as to how to distribute the two canonical length dimensions of the line element $ds^{2}=g_{\mu\nu}dx^{\mu}dx^{\nu}$ across the constituent factors.} Hence, the magnitude of the canonical length dimension of the gravitational coupling is a quantitative measure of the degree to which Weyl symmetry is broken.

The Einstein-Hilbert action of general relativity is a special case of Eq.~(\ref{a2}), namely when $a=2$ and $b=4$~\cite{Sotiriou:2008rp}. Since in this case $[\kappa]=2$, it is not locally scale-invariant. The fact that the gravitational coupling has a positive canonical length dimension can be viewed as the origin of the power-counting non-renormalizability of general relativity since it introduces a length scale into the theory. Making general relativity a locally scale-invariant theory at high energies therefore requires the canonical length dimension of the gravitational coupling to run from $[\kappa]=2$ in the IR to $[\kappa]=0$ in the UV. Since $a=2$ is fixed in general relativity, this scenario would require the spacetime dimension $b$ to run from $b=4$ in the IR to $b=2$ in the UV. Indeed, general relativity as a perturbative quantum field theory is known to be power-counting renormalizable and locally scale-invariant in two dimensions~\cite{Carlip:2017eud}.

Remarkably, several independent approaches to quantum gravity have reported just this type of dynamical dimensional reduction~\cite{Carlip:2009kf,Carlip:2017eud,Carlip:2019onx}. These approaches typically assume the leading order term is linear in the scalar curvature $R$, regardless of the particular truncation considered. That is, $a=2$ is fixed even in the extreme UV. Almost universally this results in an observed reduction of the so-called spectral dimension from $b=4$ in the IR to $b=2$ in the UV, a dimensionality for which gravity is known to become locally scale-invariant and renormalizable. 

The problem with achieving renormalizability via dimensional reduction is the unphysical consequences. For example, a reduction to $b=3$ spacetime dimensions means there are no propagating gravitational modes at all, and the Riemann tensor is identically zero implying a flat spacetime geometry~\cite{Gott86,Strobl:1999wv}. Moreover, the Einstein-Hilbert action in $b=2$ dimensions yields an Einstein tensor $G_{\mu\nu}$ that vanishes identically, and there are no field equations whatsoever~\cite{Strobl:1999wv}. Additionally, a dynamical reduction of the spacetime dimension implies that the speed of light in a vacuum must depend on its energy~\cite{Coumbe:2015zqa,Sotiriou:2011aa}. However, recent observations by the Large High Altitude Air Shower Observatory (LHAASO) find the speed of light to be linearly independent of energy up to at least $10 E_{P}$ with a $95\%$ confidence level, where $E_{P}$ is the Planck energy~\cite{LHAASO:2024lub}. LHAASO also finds the speed of light quadratically independent of energy up to at least $6\times 10^{-8} E_{P}$~\cite{LHAASO:2024lub}. Therefore, it is becoming increasingly clear from theory and experiment that dynamical dimensional reduction is not a physically viable route towards a local scale-invariant and renormalizable theory of gravity at high energies.    

We propose a more physically viable, but mathematically equivalent, alternative route. Consider fixing the spacetime dimensionality to $b=4$, and instead placing all the dynamics in the exponent $a$. In this case, the restriction to a constant dimensionality of $b=4$ coupled with the requirement that gravity become locally scale-invariant in the UV would instead force the exponent $a$ to run from $a=2$ in the IR to $a=4$ in the UV. In this way, gravity can achieve renormalizability in the UV without the unphysical consequences associated with dynamical dimensional reduction. 

We contend that the ubiquitous appearance of dynamical dimensional reduction in quantum gravity is due to the restriction of a fixed exponent $a=2$ in approaches to quantum gravity that artificially forces $b$ to vary such that the theory approaches local scale invariance. We therefore posit a mathematical equivalence between dynamical dimensional reduction in the metric formalism and AWIG in the Palatini formalism. By exploiting this equivalence, we will deduce how the exponent $a$ must vary based on the known dynamics of the mathematically equivalent, although physically inequivalent, varying spacetime dimensionality $b$.    

Since the gravitational coupling has length dimensions $[\kappa]=b-a$, if we fix $a=2$ and allow $b$ to vary then $[\kappa]_{a}=b-2$, where $[\kappa]_{a}$ refers to the dimensionality of the gravitational coupling for a fixed exponent $a$. Conversely if we fix $b=4$ and allow $a$ to vary then $[\kappa]_{b}=4-a$. We conjecture that these two statements are mathematically equivalent. That is, $[\kappa]_{a}=[\kappa]_{b}$, which yields $a=6-b$. Identifying $a/2$ with the running exponent $n(R)$ initially defined in the metric formalism for convenience, we then find $2n\left(R\right)=6-b$, and hence

\begin{equation}\label{DtoNconversion}
n\left(R\right)=3-\frac{b}{2}.
\end{equation}

Assuming the spacetime dimension $b$ can be associated with the spectral dimension $d_{s}$, which is defined via\interfootnotelinepenalty=10000 \footnote{\scriptsize For motivation behind this assumption see Refs.~\cite{Laihobb,Shomer:2007vq,Carlip:2017eud,Carlip:2009kf}.}

\begin{equation}
d_{s}=d_{t}-2\frac{\sum_{n=1}^{n=\infty} n a_{n}\sigma^{n}}{\sum_{n=0}^{n=\infty} a_{n}\sigma^{n}},
\end{equation}

\noindent where $d_{t}=4$ is the topological dimension of the manifold, $\sigma$ is the diffusion time and $a_{n}$ are a series of curvature invariants defined in the metric formalism, the first three of which are given by~\cite{Benedetti:2009ge}

\begin{equation}
a_{0}=\int_{M}d^{4}x\sqrt{-g}, \quad a_{1}=\frac{1}{6}\int_{M}d^{4}x\sqrt{-g}R, \quad a_{2}=\frac{1}{360}\int_{M}d^{4}x\sqrt{-g}\left(5R^{2}-2R_{\mu\nu}R^{\mu\nu}+2R_{\mu\nu\rho\tau}R^{\mu\nu\rho\tau}\right), 
\end{equation}

\noindent then

\begin{equation}\label{nofr1}
n\left(R\right)=1+\frac{\sum_{n=1}^{n=\infty} n a_{n}\sigma^{n}}{\sum_{n=0}^{n=\infty} a_{n}\sigma^{n}}.
\end{equation}

Expanding the sums we have $\sum_{n=1}^{n=\infty} n a_{n}\sigma^{n}=a_{1}\sigma+2a_{2}\sigma^{2}+...$ and $\sum_{n=0}^{n=\infty} a_{n}\sigma^{n}=a_{0} + a_{1}\sigma+ a_{2}\sigma^{2}+...$. All higher-order terms $a_{n}$, for which $n\geq 2$, cannot appear in the definition of $n\left(R\right)$ because they are not pure functions of the curvature scalar, and hence do not evade Ostragadski's instability~\cite{Ostrogradsky,Woodard:2015zca}. The only curvature invariants $a_{n}$ that are pure $f(R)$ functions are $a_{0}$ and $a_{1}$. Thus, we obtain

\begin{equation}\label{nofr2}
  n\left(R\right)=1+\frac{1}{1+ \frac{a_{0}}{a_{1}\sigma}}.
\end{equation}

\noindent Using the definition

\begin{equation}
R\equiv \frac{\int_{\mathcal{M}}d^{4}x\sqrt{-g}R}{\int_{\mathcal{M}}d^{4}x\sqrt{-g}}, 
\end{equation}

\noindent we then find

\begin{equation}\label{nofr3}
  n\left(R\right)= 1+\frac{1}{1+ \left(c R\right)^{-1}}\equiv 2-\frac{1}{1+c R},
\end{equation}

\noindent where $c=\sigma/6$ and has canonical length dimensions $L^{2}$, which sets a scale. We could now express our Lagrangian density as a hybrid model using the metric scalar curvature $R$ in the exponent and the Palatini scalar curvature $\mathcal{R}$ in the base, hence

\begin{equation}\label{hybrid}
f(\mathcal{R},R)=\mathcal{R}^{2-\frac{1}{1+c R}}.
\end{equation}

\noindent However, the problem with expressing the Lagrangian density in this way is that it is a function of two variables $\mathcal{R}$ and $R$, which makes it difficult to analyse numerically.  

Alternatively, it is possible to express $R$ in the metric formalism in terms of $\mathcal{R}$ in the Palatini formalism  via the expression~\cite{Sotiriou:2008rp}

\begin{equation}\label{Rconversion}
R=\mathcal{R}-\frac{3}{f'\left(\mathcal{R}\right)}\left(\frac{1}{2f'\left(\mathcal{R}\right)}\left(\nabla_{\mu}f'\left(\mathcal{R}\right)\right)\left(\nabla^{\mu}f'\left(\mathcal{R}\right)\right)+\Box f'\left(\mathcal{R}\right)\right),
\end{equation}

\noindent so that the final result is

\begin{equation}\label{nofr4}
  n\left(\mathcal{R}\right)= 2-\frac{1}{1+ c\left(\mathcal{R}-
    \frac{3}{f'\left(\mathcal{R}\right)} \left(\frac{1}{2f'\left(\mathcal{R}\right)}\left(\nabla_{\mu}f'\left(\mathcal{R}\right)\right)\left(\nabla^{\mu}f'\left(\mathcal{R}\right)\right)+\Box f'\left(\mathcal{R}\right)\right)\right)}.
\end{equation}

\noindent In this way the Lagrangian density is expressed purely in terms of the Palatini scalar curvature, which conveys some numerical advantages particularly when it comes to exploring the models cosmological dynamics~\cite{Coumbe:2021qid}.\interfootnotelinepenalty=10000 \footnote{\scriptsize Since $n\left(R\right)$ is independently defined via external physical principles, namely, dimensional reduction and Weyl invariance, and not by Eq.~(\ref{nofr4}), the construction is logically non-circular.} 


\end{section}


\begin{section}{Curing the problems with Palatini gravity}

  Palatini $f(\mathcal{R})$ gravity has several problems that must be overcome if it is to be considered a viable alternative to general relativity. Four problems commonly discussed in the literature are~\cite{Sotiriou:2008rp}

  \begin{itemize}
  \item An incorrect weak-field limit.
  \item Conflict with the Standard Model of particle physics.
  \item An ill-defined initial value problem (Cauchy problem).
  \item The appearance of unphysical surface singularities.
  \end{itemize}
  
  The origin of all four of these problems stem from the structure of the traced field equations~\cite{Sotiriou:2008rp}, namely

\begin{equation}\label{b1}
T=f'(\mathcal{R}) \mathcal{R}-2f(\mathcal{R}).
\end{equation}

\noindent To illustrate the problem with a specific example, consider the model defined by

\begin{equation}\label{b2}
f(\mathcal{R})=\mathcal{R}-\frac{\alpha^{2}}{\mathcal{R}},
\end{equation}

\noindent where $\alpha$ is a constant with appropriate dimensions~\cite{Sotiriou:2008rp,Vollick:2003aw}. Applying this specific $f(\mathcal{R})$ model to Eq.~(\ref{b1}) yields a traced stress-energy tensor

\begin{equation}\label{b3}
T=\frac{3\alpha^{2}}{\mathcal{R}}-\mathcal{R},
\end{equation}

\noindent which is a simple algebraic function of $\mathcal{R}$. One can then solve this equation for $\mathcal{R}$, resulting in a purely algebraic function of $T$ given by

\begin{equation}\label{b4}
\mathcal{R}=\frac{1}{2}\left(-T \pm \sqrt{12 \alpha^{2}+T^{2}}\right).
\end{equation}

\noindent Therefore, whenever $\mathcal{R}$ appears within the Palatini formalism it can be replaced with $F\left(T\right)$, where $F\left(T\right)$ is some algebraic function of $T$. As explained in Refs.~\cite{Sotiriou:2008rp,Barausse:2007ys} this feature is the root cause of the aforementioned problems in Palatini gravity.   

However, in metric $f(R)$ gravity there is no such problem. This is because the traced stress-energy tensor in metric $f(R)$ gravity is given by

\begin{equation}\label{b5}
T=3 \Box f'(R) + f'(R) R-2f(R).
\end{equation}

\noindent The key difference now is that $T$ is no longer a purely algebraic function of $R$, for a given function $f(R)$. Crucially, $T$ and $R$ are now related via a differential equation. We will now show that our model, despite being defined in the Palatini formalism, has a similar differential relationship between $T$ and $\mathcal{R}$, thus evading the standard problems that plague Palatini gravity. 

Plugging our specific function of $f(\mathcal{R})=\mathcal{R}^{n(\mathcal{R})}$ into Eq.~(\ref{b1}) yields

\begin{equation}\label{a6}
  T=\mathcal{R}^{n(\mathcal{R})}\left(n(\mathcal{R})-2+\mathcal{R} \rm{Log}(\mathcal{R})n'(\mathcal{R})\right).
\end{equation}

\noindent Since $n(\mathcal{R})$ given by Eq.~(\ref{nofr4}) is a differential, and not algebraic, function of $f'(\mathcal{R})$ it means that $T$ is also a differential function of $f'(\mathcal{R})$. Therefore, it is not possible to write $\mathcal{R}$ as an algebraic function of $T$, and so the aforementioned problems with the Palatini formalism are naturally avoided by this particular model.

We also highlight the fact that the Palatini formalism does not suffer from the so-called Dolgov-Kawasaki instability, regardless of the specific functional form of $f(\mathcal{R})$~\cite{Sotiriou:2006sf}. In contrast, metric $f(R)$ gravity is susceptible to the Dolgov-Kawasaki instability, unless the specific condition $f''(R)\geq 0$ is met~\cite{Sotiriou:2006sf}. Furthermore, considering AWIG as defined by Eq.~(\ref{03}), then at least for each fixed value of $n(\mathcal{R})$, we get a set of standard Palatini $\mathcal{R}^{n}$ theories, which are well-known to evade the Ostrogradsky instability problem due to their equivalence with scalar tensor theories that lead to two derivative equations of motion. Since the value of $n(\mathcal{R})$ is locally a constant at each spacetime point, this implies that at least locally AWIG evades the Ostragadski instability. However, it is important to note that globally, for non-constant $n(\mathcal{R})$, the presence of the Ostragadski instability in AWIG remains an open issue that needs further investigation. 



\end{section}


\begin{section}{A tractable approximation}

  Equation~(\ref{nofr4}) is intractable due to its complex differential form. We therefore seek an approximate tractable function for $n(\mathcal{R})$ so that we can extract a prediction from the model and test it against experimental data.

 We begin by defining $\mathcal{R_{*}}$ as the dimensionless combination $\mathcal{R_{*}}\equiv c\mathcal{R} \equiv \mathcal{R}/ \mathcal{R}_{0}$, with $\mathcal{R}_{0}$ a constant with canonical length dimensions of $[\mathcal{R}_{0}]=-2$ that represents the maximum value $\mathcal{R}$ can take. Thus, $\mathcal{R}_{*}$ is dimensionless and has a restricted domain of $[0,1]$. Moreover, by defining the dimensionless function

\begin{equation}
 \epsilon(\mathcal{R}_{*})\equiv 1 - \frac{3c}{\mathcal{R}_{*}f'\left(\mathcal{R}\right)} \left(\frac{1}{2f'\left(\mathcal{R}\right)}\left(\nabla_{\mu}f'\left(\mathcal{R}\right)\right)\left(\nabla^{\mu}f'\left(\mathcal{R}\right)\right)+\Box f'\left(\mathcal{R}\right)\right),
\end{equation}

\noindent we can rewrite Eq.~(\ref{nofr4}) as 

\begin{equation}\label{nofr5}
  n\left(\mathcal{R}_{*}\right)= 2-\frac{1}{1+ \mathcal{R}_{*}\epsilon(\mathcal{R}_{*})}.
\end{equation}

By definition, our model requires $n\left(0\right)=1$ and $n\left(1\right)=2$. We also require that the first derivative of $n\left(\mathcal{R}_{*}\right)$ with respect to $\mathcal{R}_{*}$ be zero in the IR and UV limits, that is $n'\left(0\right)=n'\left(1\right)=0$. This is because, as already shown in Ref.~\cite{Coumbe:2021qid}, we can write $n'\left(\mathcal{R}_{*}\right)$ as 

\begin{equation}\label{fp2}
  n'\left(\mathcal{R}_{*}\right)=\frac{\kappa T+\left(\mathcal{R}_{*}\mathcal{R}_{0}\right)^{n\left(\mathcal{R}_{*}\right)}\left(2-n\left(\mathcal{R}_{*}\right)\right)}{\mathcal{R}_{*}\left(\mathcal{R}_{*}\mathcal{R}_{0}\right)^{n\left(\mathcal{R}_{*}\right)}\log{\left(\mathcal{R}_{*}\mathcal{R}_{0}\right)}}.
\end{equation} 

\noindent In the UV limit we require exact Weyl invariance, which means the traced stress-energy tensor must vanish~\cite{Mukhanov:2007zz}. Thus, as $n\left(\mathcal{R}_{*}\right)\to 2$ we must also have $T \to 0$. Applying the limits $n\left(\mathcal{R}_{*}\right)\to 2$ and $T \to 0$ to Eq.(\ref{fp2}) yields $n'\left(\mathcal{R}_{*}\right)\to 0$ in the UV. Similarly, in the IR limit as $n\left(\mathcal{R}_{*}\right)\to 1$ we must have $\kappa T\to -\mathcal{R}\equiv -\left(\mathcal{R}_{*}\mathcal{R}_{0}\right)$, and so Eq.(\ref{fp2}) also yields $n'\left(\mathcal{R}_{*}\right) \to 0$ in the IR.\interfootnotelinepenalty=10000 \footnote{\scriptsize Experimental tests of general relativity also imply $n\left(\mathcal{R}_{*}\right)$ remains indistinguishable from unity over a fairly wide range of $\mathcal{R}_{*}$ values within the weak-field regime~\cite{Clifton:2005aj}, supporting the condition $n'\left(0\right) \to 0$.} These constraints must be satisfied exactly by the full expression for $n\left(\mathcal{R}_{*}\right)$.

We know that the full expression for $n\left(\mathcal{R}_{*}\right)$ must be a differential function of $f'(\mathcal{R})$, but an approximate tractable function can be obtained by assuming that $n\left(\mathcal{R}_{*}\right)$ admits a series expansion in $\mathcal{R}_{*}$ of the form

\begin{equation}\label{a3}
n\left(\mathcal{R}_{*}\right)= \sum_{m=0}^{\infty}b_{m}\mathcal{R}_{*}^{m},
\end{equation}

\noindent where $b_{m}$ are dimensionless constants. As shown in Ref.~\cite{Coumbe:2021qid}, the lowest order polynomial to satisfy the conditions $n(0)=1$, $n(1)=2$ and $n'\left(0\right)=n'\left(1\right)=0$ is 

\begin{equation}\label{a5}
n\left(\mathcal{R}_{*}\right)=1+3\mathcal{R}_{*}^{2}-2\mathcal{R}_{*}^{3}.
\end{equation} 

\noindent Equating Eqs.~(\ref{nofr5}) and~(\ref{a5}) and solving for $\epsilon\left(\mathcal{R}_{*}\right)$ then yields

\begin{equation}\label{a7}
\epsilon\left(\mathcal{R}_{*}\right)=\frac{\mathcal{R}_{*}\left(3-2\mathcal{R}_{*}\right)}{\left(\mathcal{R}_{*}-1\right)^{2}\left(1+2\mathcal{R}_{*}\right)}.
\end{equation} 

For the particular ansatz of Eq.~(\ref{a5}) the functions $n(\mathcal{R}_{*})$, $f(\mathcal{R}_{*})$, its first derivative $f'(\mathcal{R}_{*})$ and the traced stress-energy tensor $T\left(\mathcal{R}_{*}\right)$ are plotted in Fig.~\ref{nfandfprime}.

\begin{figure}[H]
  \centering
\subfloat[\label{figa}]{\includegraphics[width=6cm]{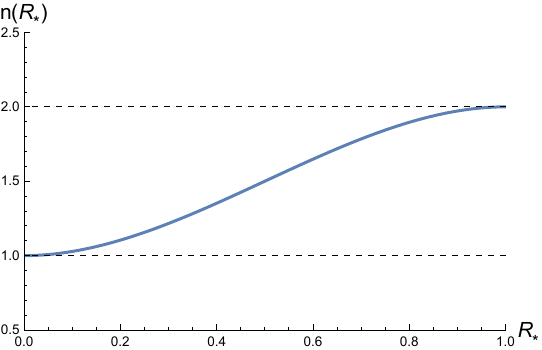} \label{fig:ABa}}
\subfloat[\label{figb}]{\includegraphics[width=6cm]{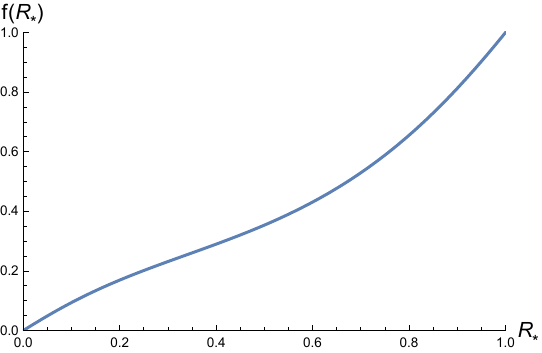} \label{fig:ABb}}\\
\subfloat[\label{figc}]{\includegraphics[width=6cm]{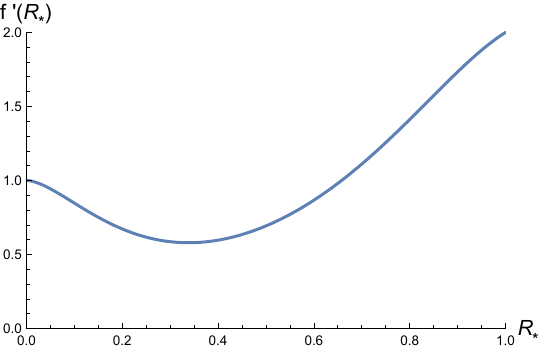} \label{fig:ABc}}
\subfloat[\label{figb}]{\includegraphics[width=6cm]{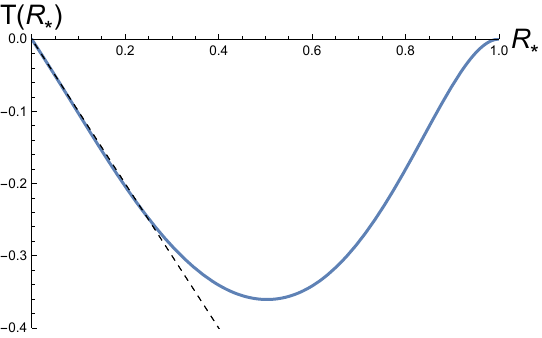} \label{fig:ABd}}
\caption{(a) The dimensionless exponent $n\left(\mathcal{R}_{*}\right)$, (b) the Lagrangian density $f\left(\mathcal{R}_{*}\right)$, (c) its first derivative $f'\left(\mathcal{R}_{*}\right)$ and (d) the traced stress-energy tensor $T\left(\mathcal{R}_{*}\right)$. The upper (lower) dashed line in (a) gives the asymptotic value in the high-curvature (low-curvature) limit. The dashed line in (d) indicates the general relativistic dependency $T\left(\mathcal{R}_{*}\right)\propto -\mathcal{R}_{*}$.}
\label{nfandfprime}
\end{figure}

\noindent In Ref.~\cite{Coumbe:2021qid}, AWIG under this particular ansatz was shown to be consistent with known cosmological dynamics. For example, this model exhibits the expected sequence of an accelerating de Sitter-like phase at early times, followed by intermediate deccelerating radiation and matter-dominated phases, before returning to an accelerating de Sitter-like phase at late times.

\end{section}


\begin{section}{Testing AWIG using binary black hole mergers}

  General relativity has passed every experimental test it has ever faced. Yet, until recently all such tests of general relativity have been in the weak-field regime. Recent observations of binary black hole mergers, however, have allowed the first-ever strong-field tests of general relativity~\cite{LIGOScientific:2016lio}. Since 2015, on the order of 100 gravitational wave events have been observed over a wide range of parameters, providing a new and data-rich testing ground for theories of gravitation.

 Assuming the gravitational field for inspiralling black hole binaries becomes sufficiently strong to probe the intermediate and high curvature regimes of our model then, as indicated by Fig.~\ref{fig:ABc}, we should expect to see a rescaling of the metric tensor by a minimum factor of $f'\left(\mathcal{R}_{*}\right)_{min}=0.58$ at some point during the inspiral phase followed by a rescaling of at most $f'\left(\mathcal{R}_{*}\right)_{max}=2$ presumably during the final stages of the merger where the scalar curvature is largest.

Now consider two stationary observers $A$ and $B$ in a spherically symmetric gravitational field described by the Schwarzschild metric. Since the observers are stationary there is no variation in the spatial Schwarzschild coordinates, hence $dr=d\theta=d\phi=0$. Let $\omega_{A}$ be the frequency according to observer $A$ and $\omega_{B}$ be the frequency according to observer $B$. In general relativity, the ratio of these frequencies is related to the ratio of the metric tensor at the location of the two observers, namely

\begin{equation}
  \frac{\omega_{B}}{\omega_{A}}=\sqrt{\frac{g_{00}(r_{A})}{g_{00}(r_{B})}},
\end{equation}

\noindent where $g_{00}$ is the temporal component of the metric tensor.

Considering the same scenario but with the rescaled metric $\tilde{g}_{\mu\nu}$ in AWIG, which is related to the standard metric $g_{\mu\nu}$ by

\begin{equation}\label{metrictrans}
  \tilde{g}_{\mu\nu}=f'(\mathcal{R})g_{\mu\nu},
\end{equation}

\noindent we find

\begin{equation}
\frac{\tilde{\omega}_{B}}{\tilde{\omega}_{A}}=\sqrt{\frac{\tilde{g}_{00}(r_{A})}{\tilde{g}_{00}(r_{B})}}=\sqrt{\frac{f'(\mathcal{R})_{A}\ g_{00}(r_{A})}{f'(\mathcal{R})_{B}\ g_{00}(r_{B})}}=\sqrt{\frac{f'(\mathcal{R})_{A}}{f'(\mathcal{R})_{B}}}\frac{\omega_{B}}{\omega_{A}}.
\end{equation}

\noindent Now, let observer A be close to the massive object and let observer $B$ be arbitrarily far from it. At an arbitrarily large distance from a massive object, we should have $f'(\mathcal{R})_{B}\to 1$, which implies $\tilde{\omega}_{B}=\omega_{B}$. However, for observer $A$ who is close to the massive object $f'(\mathcal{R})_{A}\neq 1$. Therefore, in this case, we arrive at the result

\begin{equation}
\frac{\omega_{A}}{\tilde{\omega}_{A}}=\sqrt{f'(\mathcal{R})_{A}}.
\end{equation}

Since $f'(\mathcal{R})$ very close to the instant of a black hole merger is expected to approach a maximum value of $2$, we predict an upper bound of $\omega_{GR}/ \omega_{AWIG}=\sqrt{2}$. That is, the observed frequency of gravitational waves emitted at the instant the two black holes merge should be lower than the general relativistic prediction by a factor of at most $1.41$. At slightly lower curvature scales, presumably during the inspiral phase, we predict a lower bound of $\omega_{GR}/ \omega_{AWIG}\simeq \sqrt{0.58} \simeq 0.76$. Therefore, we predict deviations below unity to be bound by

\begin{equation}
0.76 \lesssim \left(\frac{\omega_{GR}}{\omega_{AWIG}}\right)_{min}\lesssim 1,
\end{equation}

\noindent and deviations above unity to be bound by

\begin{equation}
1 \lesssim \left(\frac{\omega_{GR}}{\omega_{AWIG}}\right)_{max} \lesssim 1.41.
\end{equation}

Thus, a statistically significant deviation above unity but less than or equal to $(\omega_{GR}/ \omega_{AWIG})_{max}=1.41$ during the final merge phase when curvature is highest would favour AWIG over GR. Likewise, a statistically significant deviation below unity but greater than or equal to $(\omega_{GR}/ \omega_{AWIG})_{min}\simeq 0.76$ at lower curvatures would also favour AWIG over GR. Crucially, during the inspiral the dip below unity should occur before a subsequent rise above unity just before the black holes coalesce. We also point out that the gravitational field of the merging black holes may not be strong enough to probe the high curvature limit of our theory and so the predicted deviation $(\omega_{GR}/ \omega_{AWIG})_{max}$ is an upper bound only.~\interfootnotelinepenalty=10000 \footnote{\scriptsize In addition to the asatz of Eq.~(\ref{a5}), we have checked the robustness of this prediction against other functions for $n\left(\mathcal{R}_{*}\right)$, namely a trigonometric function $n\left(\mathcal{R}_{*}\right)=(1/2)(3-\rm{cos}(\pi \mathcal{R}_{*}))$ and a logistic function $n\left(\mathcal{R}_{*}\right)=1+\frac{1}{1+ e^{-k(\mathcal{R}_{*}-d)}}$. In both cases we find $2 \leq f'(\mathcal{R}_{*})_{max}\leq 2.046$ and $0.570 \leq f'(\mathcal{R}_{*})_{min}\leq 0.901$. See appendix~\ref{appx1} for more details.} Since the scalar curvature close to the event horizon of smaller black holes is greater than that of larger black holes we would also expect $\omega_{GR}/ \omega_{AWIG}$ to grow as we study binary black holes of lower total mass during the final merge phase.

\begin{subsection}{GW150914}

The gravitational-wave event GW150914 was independently detected by the Laser Interferometer Gravitational-wave Observatory (LIGO) in Hanford, Washington and by LIGO Livingston in Louisiana on 14 September, 2015 at 09:50:45 UTC with a signal-to-noise ratio of 24 and a significance of 5.1 sigma.~\cite{LIGOScientific:2016aoc}. GW150914 was most likely caused by the inward spiral and subsequent merger of a pair of black holes of respective masses $36^{+5}_{-4}$ and $29^{+4}_{-4}$ solar masses~\cite{LIGOScientific:2016aoc}.  

General relativistic predictions specific to the black holes thought to cause GW150914 have been computed using intensive numerical simulations of general relativity~\cite{LIGOScientific:2016aoc,Campanelli:2005dd,Blanchet:2009sd}. For example, the gravitational wave strain defined by $h(t)=\Delta L(t)\L$, where $\Delta L(t)=\delta L_{x}-\delta L_{y}$ is the time-dependent change in length of two perpendicular interferometer arm lengths $L_{x}$ and $L_{y}$, has been computed via extremely precise numerical relativity simulations~\cite{LIGOScientific:2016aoc,LIGOScientific:2019lzm}. Note that all time series presented in this section have been filtered with a 35–350 Hz band-pass filter to suppress large fluctuations outside the detectors’ most sensitive frequency band, and band-reject filters to remove the strong instrumental spectral lines~\cite{LIGOScientific:2016aoc}.

\begin{subsubsection}{Hanford}\label{H1}

Simulations of $h(t)$ for GW150914 are projected onto the Hanford observatory, with the result shown in Fig.~\ref{fig:8ABa}~\cite{LIGOScientific:2016aoc,LIGOScientific:2019lzm}. From the data used to make Fig.~\ref{fig:8ABa}, we can determine the frequency of the gravitational waves as a function of the inspiral time. To do this we find the time between adjacent discrete time values $t_{i}$ and $t_{i+1}$ and calculate the frequency via $\omega(t)=(4n(t_{i+1}-t_{i}))^{-1}$, where $n$ is the number of adjacent pairs in each $1/4$ of a cycle.\interfootnotelinepenalty=10000 \footnote{\scriptsize In addition to this method, the frequency $\omega(t)$ has also been computed using another method to check for consistency. See Appendix~\ref{appendix2} for details.} The function $\omega(t)$ is then smoothed out using a third-order interpolation function. Figure~\ref{fig:8ABb} shows the resulting $\omega(t)$ plot including error bands representing one standard deviation taken over three adjacent $\omega(t)$.  

\begin{figure}[H]
  \centering
  \subfloat[\label{figa}]{\includegraphics[width=9cm]{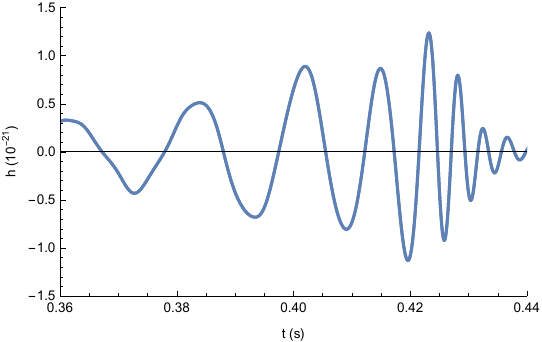} \label{fig:8ABa}}
  \subfloat[\label{figb}]{\includegraphics[width=8.89cm]{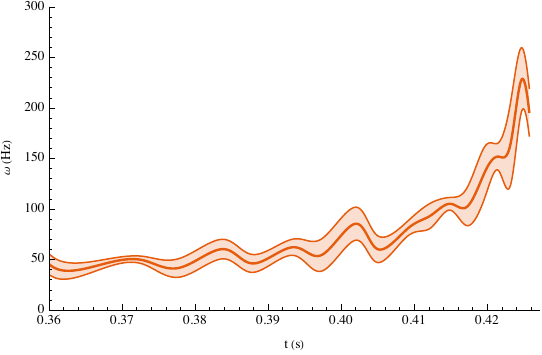} \label{fig:8ABb}}
  \caption{(a) Strain versus time and (b) frequency versus time from a numerical relativity simulation of GW150914 including a band-pass filter~\cite{LIGOScientific:2016aoc}. Error bands in (b) represent one standard deviation taken over three adjacent $\omega(t)$ values.} \label{fig:AB}
\end{figure}

We now seek the corresponding $\omega(t)$ values observed by LIGO Hanford. We perform a pixel analysis of the spectrogram data released by the LIGO collaboration shown in Fig.~\ref{fig:9ABa}~\cite{LIGOScientific:2016aoc}. An analysis of the pixel cross-section for 11 different times is performed. An example pixel cross-section at $t=0.425$ is shown in Fig.~\ref{fig:9ABb}. The error in the maximum pixel intensity is one standard deviation from the peak as calculated using the relationship between the full width at half maximum (FWHM) and the standard deviation $\sigma$ given by $FWHM=2\sqrt{2\rm{ln}(2)}\sigma\approx 2.355 \sigma$~\cite{Mizutani:2017ggu}. The FWHM for $t=0.425$ is depicted in Fig.~\ref{fig:9ABb}. Pixel numbers are converted to $\omega(t)$ values via the scale in Fig.~\ref{fig:9ABa}. The gravitational wave frequency values predicted by GR for  GW150914 can now be compared with those observed by LIGO Hanford via the ratio $\omega_{GR}/\omega_{Han}$. The results are summarized in Fig.~\ref{Money1} and Tab.~\ref{tab2}.         

\begin{figure}[H]
  \centering
    \subfloat[\label{figa}]{\includegraphics[width=9cm]{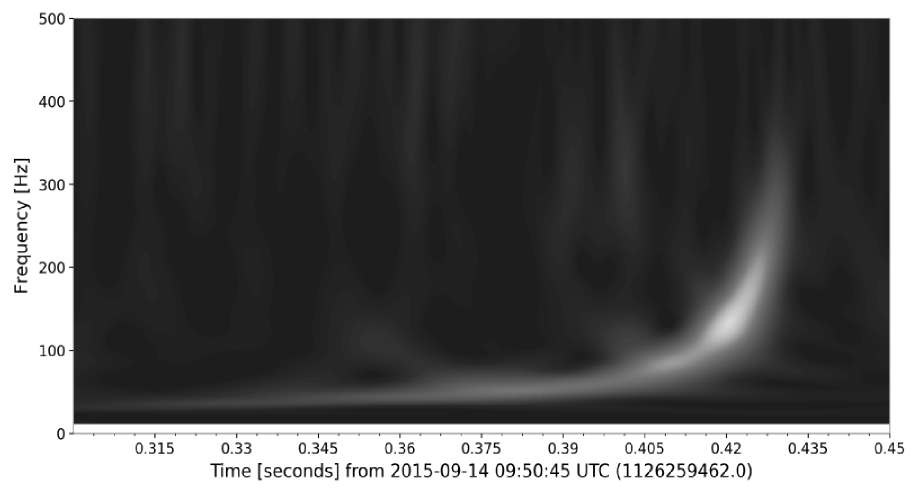} \label{fig:9ABa}}
  \subfloat[\label{figb}]{\includegraphics[width=7.1cm]{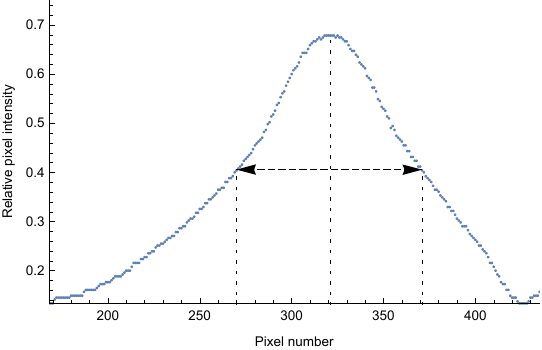} \label{fig:9ABb}}
  \caption{(a) Frequency versus time for the gravitational-wave event GW150914 observed by LIGO Hanford. (b) An example pixel cross-section at $t=0.425$, where the dashed vertical lines show the peak intensity and the FWHM.} \label{fig:AB}
\end{figure}

As can be seen in Fig.~\ref{Money1}, the ratio $\omega_{GR}/\omega_{Han}$ at $t\leq 0.365 s$ relatively far from the merge point is consistent with the GR prediction of unity. For intermediate times during the inspiral phase ($0.365 s \leq t \leq 0.41 s$) there appears to be a small but consistent dip below unity, dropping as low as $0.8$. In the merger phase ($t\geq 0.41 s$) there is a slight rise above unity, climbing as high as $1.3$. Although these results have low statistical significance, the overall behaviour is consistent with what AWIG predicts.

\begin{figure}[H]
\centering
\scalebox{1.0}{\includegraphics{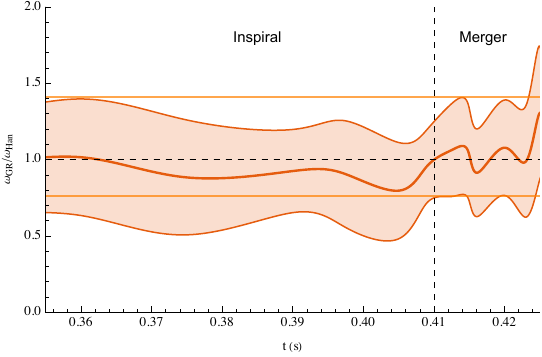}}
\caption{The ratio $\omega_{GR}/\omega_{Han}$ as a function of time including error bands. The dashed horizontal line indicates the general relativistic prediction and the dashed vertical line indicates the approximate boundary between the inspiral and merge phases~\cite{LIGOScientific:2016aoc}. The upper shaded region indicates the AWIG prediction $(\omega_{GR}/ \omega_{AWIG})_{max}$ and the lower shaded region indicates the AWIG prediction $(\omega_{GR}/ \omega_{AWIG})_{min}$.}
\label{Money1}
\end{figure}

\begin{table}[H]
\begin{center}
\begin{tabular} {|c||c|c|c|c|}
\hline
t(s) & $\omega_{GR}$ & $\omega_{Han}$ & $\omega_{GR}$/$\omega_{Han}$ & $\sigma$ tension with $\omega_{GR}$\\ \hline
\hline
0.36000 & $45.2 \pm 10.7$ & $44.6 \pm 13.0$ & $1.0 \pm 0.4$ & $0.0$ \\ \hline
0.37500 & $44.1 \pm 7.5$ & $50.0 \pm 19.6$  & $0.9 \pm 0.4$ & $0.3$ \\ \hline
0.39000 & $51.2 \pm 7.6$ & $55.4 \pm 14.1$ & $0.9 \pm 0.3$ & $0.3$ \\ \hline
0.39750 & $55.0 \pm 16.2$ & $60.9 \pm 15.2$ & $0.9 \pm 0.3$ & $0.3$ \\ \hline
0.40500 & $60.5 \pm 13.5$ & $76.1 \pm 25.0$ & $0.8 \pm 0.3$ & $0.5$ \\ \hline
0.40875 & $79.8 \pm 7.4$ & $83.7\pm 19.6$  & $1.0 \pm 0.2$ & $0.2$ \\ \hline
0.41250 & $94.5 \pm 12.1$ & $89.1\pm 22.8$  & $1.1 \pm 0.3$ & $0.2$ \\ \hline
0.41625 & $101.3 \pm 11.2$ & $110.9\pm 32.6$  & $0.9 \pm 0.3$ & $0.3$ \\ \hline
0.42000 & $140.4 \pm 24.2$ & $130.4\pm 30.4$  & $1.1 \pm 0.3$ & $0.3$ \\ \hline
0.42310 & $161.3 \pm 38.0$ & $163.0 \pm 44.6$ & $1.0 \pm 0.4$ & $0.0$ \\ \hline
0.42500 & $228.1 \pm 29.6$ & $175.0 \pm 54.3$ & $1.3 \pm 0.4$  & $0.9$ \\ \hline
\end{tabular}
\end{center}
\caption{A table showing the 11 times at which the frequency predicted by general relativity $\omega_{GR}$ is compared with the frequency measured by the Hanford observatory $\omega_{Han}$ for GW150914. The ratio $\omega_{GR}$/$\omega_{Han}$ is given in column four and the tension between $\omega_{Han}$ and the general relativistic prediction $\omega_{GR}$ is given in column five.}  
\label{tab2}
\end{table}

As shown in Tab.~\ref{tab2}, at $t=0.42500 s$ there is a $0.9\ \sigma$ tension between the observed value $\omega_{Han}$ and the GR prediction $\omega_{GR}$, whereas at the same point, there is only a $0.2\ \sigma$ tension between the observed value $\omega_{Han}$ and the AWIG prediction $\omega_{AWIG}$.\interfootnotelinepenalty=10000 \footnote{\scriptsize This comparison assumes the uncertainty in $\omega_{AWIG}$ is the same as for $\omega_{GR}$. We also assume that $(\omega_{GR}/ \omega_{AWIG})_{max}=1.41$ only applies to the last time stamp before the merge, while $(\omega_{GR}/ \omega_{AWIG})_{min}=0.76$ applies to points during the inspiral phase.} At $t=0.405 s$, there is a $0.5\ \sigma$ tension between the observed value $\omega_{Han}$ and the GR prediction $\omega_{GR}$ but only a $0.1\ \sigma$ tension between the observed value $\omega_{Han}$ and the AWIG prediction $\omega_{AWIG}$. At least for these two points, the data therefore marginally favours AWIG over GR. 

\end{subsubsection}
\begin{subsubsection}{Livingston}

The gravitational wave strain predicted by general relativity for GW150914 and projected on to the Livingston observatory (L1) can be seen in Fig.~\ref{fig:11ABa}. The resulting $\omega(t)$ plot is shown in Fig.~\ref{fig:11ABb}. The methodology employed in this subsection is identical to that already detailed in subsection~\ref{H1}, and so is not repeated here. Results for our analysis of the Livingston detector are presented in Fig.~\ref{Money2} and Tab.~\ref{tab3}. 
  
\begin{figure}[H]
  \centering
    \subfloat[\label{figa}]{\includegraphics[width=9cm]{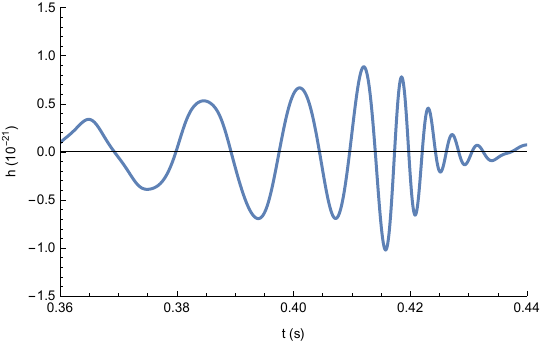} \label{fig:11ABa}}
  \subfloat[\label{figb}]{\includegraphics[width=8.89cm]{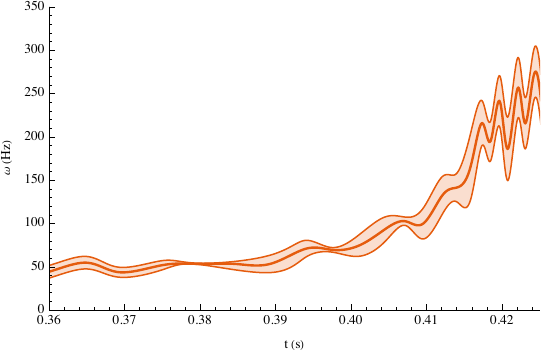} \label{fig:11ABb}}
  \caption{(a) Strain versus time and (b) frequency versus time from a numerical relativity simulation of GW150914 projected on to L1~\cite{LIGOScientific:2016aoc}. Error bands represent one standard deviation in the frequency taken over three adjacent $\omega(t)$ values.} \label{fig:AB}
\end{figure}

Figure~\ref{Money2} shows that the ratio $\omega_{GR}/\omega_{Liv}$ far from the merger at $t\leq 0.38 s$ is consistent with the GR prediction of unity. For intermediate times during the inspiral phase ($0.38 s \leq t \leq 0.41 s$) there is a slight but consistent dip below unity, going as low as $0.8$. For points in the merger phase ($t\geq 0.41 s$) there is a slight rise above unity, going as high as $1.3$ at two points. Again, this is exactly the general behaviour we would expect based on the AWIG model, although these results also have low statistical significance due to the large error bars.   

\begin{figure}[H]
\centering
\scalebox{1.0}{\includegraphics{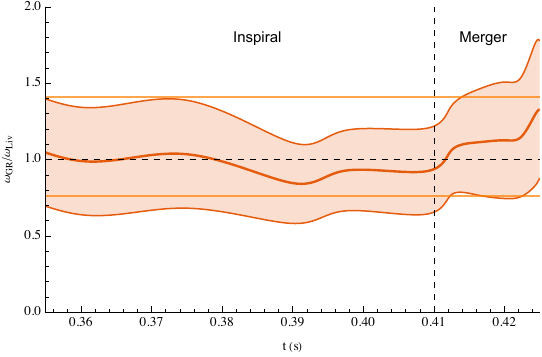}}
\caption{The ratio $\omega_{GR}/\omega_{Liv}$ as a function of time. The dashed horizontal line indicates the general relativistic prediction and the dashed vertical line indicates the approximate boundary between the inspiral and merge phases~\cite{LIGOScientific:2016aoc}. The upper shaded region indicates the AWIG prediction $(\omega_{GR}/ \omega_{AWIG})_{max}$ and the lower shaded region indicates the AWIG prediction $(\omega_{GR}/ \omega_{AWIG})_{min}$.}
\label{Money2}
\end{figure}

\begin{table}[H]
\begin{center}
\begin{tabular} {|c||c|c|c|c|}
\hline
t(s) & $\omega_{GR}$ & $\omega_{Liv}$ & $\omega_{GR}$/$\omega_{Liv}$ & $\sigma$ tension with $\omega_{GR}$\\ \hline
\hline
0.36000 & $44.1 \pm 7.1$ & $44.6 \pm 14.1$ & $1.0 \pm 0.4$ & $0.0$ \\ \hline
0.37500 & $51.7 \pm 5.3$ & $50.0 \pm 16.3$  & $1.0 \pm 0.4$ & $0.1$ \\ \hline
0.39000 & $55.1 \pm 11.5$ & $65.2 \pm 15.2$ & $0.8 \pm 0.3$ & $0.5$ \\ \hline
0.39750 & $69.5 \pm 5.0$ & $75.0 \pm 20.7$ & $0.9 \pm 0.3$ & $0.3$ \\ \hline
0.40875 & $98.1 \pm 12.9$ & $106.5 \pm 29.3$ & $0.9 \pm 0.3$ & $0.3$ \\ \hline
0.41250 & $137.0 \pm 19.3$ & $129.3\pm 30.4$  & $1.1 \pm 0.3$ & $0.2$ \\ \hline
0.41625 & $181.3 \pm 38.6$ & $163.0\pm 37.0$  & $1.1 \pm 0.3$ & $0.3$ \\ \hline
0.42000 & $228.7 \pm 33.9$ & $203.3\pm 62.0$  & $1.1 \pm 0.4$ & $0.4$ \\ \hline
0.42240 & $252.1 \pm 31.6$ & $219.6 \pm 69.6$ & $1.1 \pm 0.4$ & $0.4$ \\ \hline
0.42430 & $273.1 \pm 29.5$ & $212.0 \pm 72.8$ & $1.3 \pm 0.5$  & $0.8$ \\ \hline
0.42500 & $262.4 \pm 25.1$ & $197.8 \pm 64.1$ & $1.3 \pm 0.4$  & $0.9$ \\ \hline
\end{tabular}
\end{center}
\caption{A table showing the 11 times at which the frequency predicted by general relativity $\omega_{GR}$ is compared with the frequency measured by the Livingston observatory $\omega_{Liv}$. The ratio $\omega_{GR}$/$\omega_{Liv}$ is given in column four and the tension between $\omega_{Liv}$ and the general relativistic prediction $\omega_{GR}$ is given in column five.}
\label{tab3}
\end{table}

As shown in Tab.~\ref{tab3}, at $t=0.42500 s$ there is a $0.9\ \sigma$ tension between the observed value $\omega_{Liv}$ and the GR prediction $\omega_{GR}$, whereas there is only a $0.2\ \sigma$ tension between the observed value $\omega_{Liv}$ and the AWIG prediction $\omega_{AWIG}$ at the same point. At $t=0.39000 s$, there is a $0.5\ \sigma$ tension between the observed value $\omega_{Liv}$ and the GR prediction $\omega_{GR}$, but a $0.4\ \sigma$ tension between the observed value $\omega_{Liv}$ and the AWIG prediction $\omega_{AWIG}$. At least for these two points, the data again appears to marginally favour AWIG over GR.

\end{subsubsection}
\end{subsection}
\begin{subsection}{GW151226}

GW151226 was most likely caused by the inward spiral and subsequent merger of a pair of black-holes of respective masses $14.2^{+8.3}_{-3.7}$ and $7.5^{+2.3}_{-2.3}$ solar masses~\cite{LIGOScientific:2016sjg}. The event occurred at GPS time 1135136350.65 = December 26 2015, 03:38:53.65 UTC and was recovered with a network signal-to-noise ratio of 13 and a significance of greater than 5 sigma~\cite{LIGOScientific:2016sjg}. 

\begin{subsubsection}{Hanford}

The gravitational wave strain $h(t)$ computed via full numerical relativity simulations for parameters specific to GW151226 and without any filtering is shown in Fig.~\ref{fig:12ABa}~\cite{LIGOScientific:2016sjg,LIGOScientific:2019lzm}.~\interfootnotelinepenalty=10000 \footnote{\scriptsize Unfortunately, simulations for GW151226 that include a band-pass filter were not available to the authors, although based on Appendix~\ref{appendix2} the filter does not seem to have a significant impact anyhow. Here, GWOSC simulations~\cite{GWOSC} without a band-pass filter are used for an observer at spatial infinity, at extrapolation order $N=4$ and using the dominant mode $l = 2$, $m=2$~\cite{LIGOScientific:2016sjg,LIGOScientific:2019lzm}. The same numerical relativity data is used in Section~\ref{livgw151226}.} 
  
\begin{figure}[H]
  \centering
    \subfloat[\label{figa}]{\includegraphics[width=9cm]{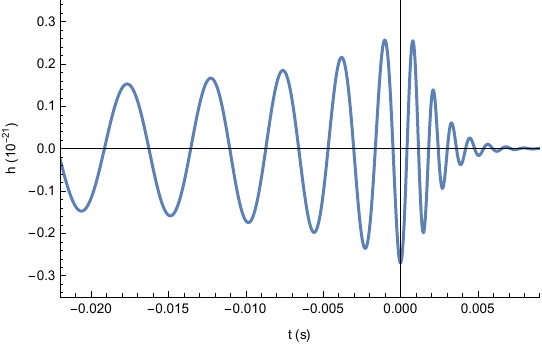} \label{fig:12ABa}}
  \subfloat[\label{figb}]{\includegraphics[width=8.89cm]{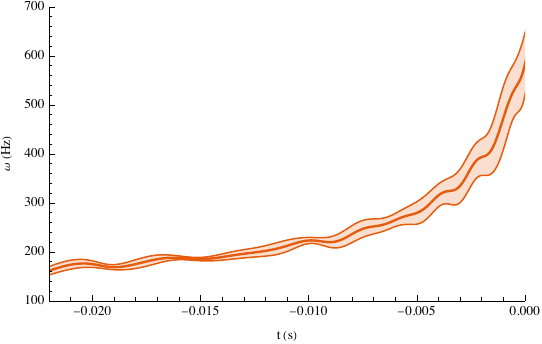} \label{fig:12ABb}}
  \caption{(a) Strain versus time and (b) frequency versus time from a numerical relativity simulation of GW151226~\cite{LIGOScientific:2016sjg}. Error bands represent one standard deviation in the frequency taken over three adjacent $\omega(t)$ values.} \label{fig:GW151226Han}
\end{figure}

 Results for our analysis of the Hanford observation of GW151226 are presented in Fig.~\ref{Money3} and Tab.~\ref{tab4}. Figure~\ref{Money3} shows that the ratio $\omega_{GR}/\omega_{Han}$ for $t\leq -0.007 s$ is consistent with the GR prediction of unity. For $-0.007 s \leq t \leq -0.001 s$ there is a statistically significant dip below unity, reaching as low as $0.8$. Close to the merge point ($t= 0 s$), there is a statistically significant rise above unity to a value of $1.3$. Again, this is consistent with the general behaviour predicted by AWIG. 

\begin{figure}[H]
\centering
\scalebox{1.0}{\includegraphics{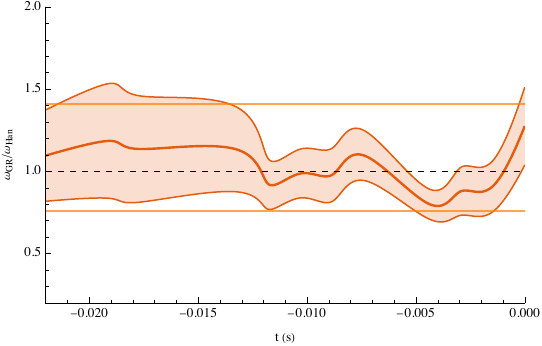}}
\caption{The ratio $\omega_{GR}/\omega_{Han}$ as a function of time for GW151226. The dashed horizontal line indicates the general relativistic prediction. The upper shaded region indicates the AWIG prediction $(\omega_{GR}/ \omega_{AWIG})_{max}$ and the lower shaded region indicates the AWIG prediction $(\omega_{GR}/ \omega_{AWIG})_{min}$.}
\label{Money3}
\end{figure}

\begin{table}[H]
\begin{center}
\begin{tabular} {|c||c|c|c|c|}
\hline
t(s) & $\omega_{GR}$ & $\omega_{Han}$ & $\omega_{GR}$/$\omega_{Han}$ & $\sigma$ tension with $\omega_{GR}$\\ \hline
\hline
-0.01912 & $171.0 \pm 3.0$ & $144.4 \pm 42.4$ & $1.2 \pm 0.3$ & $0.6$ \\ \hline
-0.01765 & $172.9 \pm 12.9$ & $152.3 \pm 41.8$ & $1.1 \pm 0.3$ & $0.5$ \\ \hline
-0.01324 & $195.5 \pm 13.0$ & $172.5 \pm 37.5$  & $1.1 \pm 0.3$ & $0.6$ \\ \hline
-0.01176 & $204.0 \pm 19.0$ & $221.5 \pm 30.4$ & $0.9 \pm 0.2$ & $0.5$ \\ \hline
-0.01029 & $220.5 \pm 8.0$ & $223.5 \pm 32.6$ & $1.0 \pm 0.1$ & $0.1$ \\ \hline
-0.00882 & $221.0 \pm 13.0$ & $225.5 \pm 34.9$ & $1.0 \pm 0.2$ & $0.1$ \\ \hline
-0.00735 & $249.0 \pm 17.0$ & $227.5\pm 27.4$  & $1.1 \pm 0.2$ & $0.7$ \\ \hline
-0.00441 & $298.0 \pm 24.5$ & $371.4\pm 29.2$  & $0.8 \pm 0.1$ & $1.9$ \\ \hline
-0.00294 & $341.0 \pm 35.5$ & $388.4\pm 50.0$  & $0.9 \pm 0.1$ & $0.8$ \\ \hline
-0.00147 & $414.5 \pm 53.5$ & $456.0 \pm 51.0$ & $0.9 \pm 0.2$ & $0.6$ \\ \hline
0.00000 & $581.4 \pm 77.6$ & $460.0 \pm 55.3$ & $1.3 \pm 0.2$  & $1.3$ \\ \hline
\end{tabular}
\end{center}
\caption{A table showing the 11 times at which the frequency predicted by general relativity $\omega_{GR}$ is compared with the frequency measured by the Hanford observatory $\omega_{Han}$. The ratio $\omega_{GR}$/$\omega_{Han}$ is given in column four and the tension between $\omega_{Han}$ and the general relativistic prediction $\omega_{GR}$ is given in column five.}
\label{tab4}
\end{table}

As can be seen in Tab.~\ref{tab4}, at $t=0 s$ there is a $1.3\ \sigma$ tension between $\omega_{Han}$ and $\omega_{GR}$, while there is only a $0.5\ \sigma$ tension between $\omega_{Han}$ and $\omega_{AWIG}$ at the same time value. At $t=-0.00441 s$, there is a $1.9\ \sigma$ tension between $\omega_{Han}$ and $\omega_{GR}$, while at the same time value there is only a $0.5\ \sigma$ tension between $\omega_{Han}$ and $\omega_{AWIG}$. Therefore, these data points are once again more consistent with AWIG than GR.

\end{subsubsection}

\begin{subsubsection}{Livingston}\label{livgw151226}

 Results for our analysis of the Livingston observation of GW151226 are summarized in Fig.~\ref{Money4} and Tab.~\ref{tab5}. Figure~\ref{Money4} shows that the ratio $\omega_{GR}/\omega_{Liv}$ for $t\leq -0.011 s$ is consistent with the GR prediction of unity. For intermediate times $-0.011 s \leq t \leq -0.001 s$ there is a statistically significant dip below unity, consistently reaching as low as $0.7$. Close to the merge point ($t= 0 s$) there is a statistically significant rise above unity reaching a value of $1.2$. Again, this also matches the general behaviour AWIG predicts. 
  
\begin{figure}[H]
\centering
\scalebox{1.0}{\includegraphics{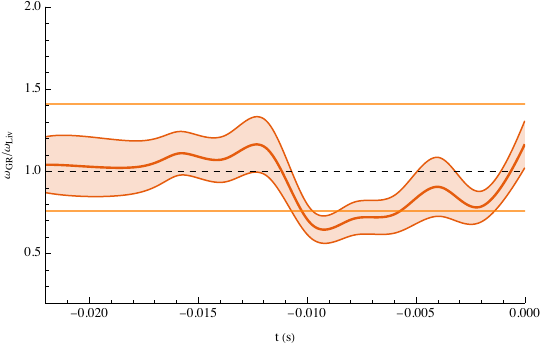}}
\caption{The ratio $\omega_{GR}/\omega_{Liv}$ as a function of time for GW151226. The dashed horizontal line indicates the general relativistic prediction. The upper shaded region indicates the AWIG prediction $(\omega_{GR}/ \omega_{AWIG})_{max}$ and the lower shaded region indicates the AWIG prediction $(\omega_{GR}/ \omega_{AWIG})_{min}$.}
\label{Money4}
\end{figure}

\begin{table}[H]
\begin{center}
\begin{tabular} {|c||c|c|c|c|}
\hline
t(s) & $\omega_{GR}$ & $\omega_{Liv}$ & $\omega_{GR}$/$\omega_{Liv}$ & $\sigma$ tension with $\omega_{GR}$\\ \hline
\hline
-0.02167 & $166.0 \pm 17.0$ & $159.7 \pm 21.6$ & $1.0 \pm 0.2$ & $0.2$ \\ \hline
-0.01773 & $171.0 \pm 14.0$ & $166.5 \pm 21.6$ & $1.0 \pm 0.2$ & $0.2$ \\ \hline
-0.01576 & $191.5 \pm 3.0$ & $172.6 \pm 20.4$  & $1.1 \pm 0.1$ & $0.9$ \\ \hline
-0.01379 & $191.0 \pm 8.0$ & $176.7 \pm 21.9$ & $1.1 \pm 0.1$ & $0.6$ \\ \hline
-0.01182 & $203.5 \pm 10.5$ & $178.9 \pm 24.2$ & $1.1 \pm 0.2$ & $0.9$ \\ \hline
-0.00985 & $223.0 \pm 16.0$ & $325.1 \pm 34.7$ & $0.7 \pm 0.1$ & $2.7$ \\ \hline
-0.00788 & $238.0 \pm 18.0$ & $335.0\pm 39.6$  & $0.7 \pm 0.1$ & $2.2$ \\ \hline
-0.00591 & $266.0 \pm 15.0$ & $357.7\pm 52.5$  & $0.7 \pm 0.1$ & $1.7$ \\ \hline
-0.00394 & $318.0 \pm 24.0$ & $351.4\pm 64.1$  & $0.9 \pm 0.2$ & $0.5$ \\ \hline
-0.00197 & $393.5 \pm 40.5$ & $500.0 \pm 28.7$ & $0.8 \pm 0.1$ & $2.1$ \\ \hline
0.00000 & $582.0 \pm 63.0$ & $500.0 \pm 28.7$ & $1.2 \pm 0.1$  & $1.2$ \\ \hline
\end{tabular}
\end{center}
\caption{A table showing the 11 times at which the frequency predicted by general relativity $\omega_{GR}$ is compared with the frequency measured by the Livingston observatory $\omega_{Liv}$ for GW151226. The ratio $\omega_{GR}$/$\omega_{Liv}$ is given in column four and the tension between $\omega_{Liv}$ and the general relativistic prediction $\omega_{GR}$ is given in column five.}
\label{tab5}
\end{table}
  
As can be seen in Tab.~\ref{tab5}, at $t=0 s$ there is a $1.2\ \sigma$ tension between $\omega_{Liv}$ and $\omega_{GR}$, however there is a slightly lower tension of $1.0\ \sigma$ between $\omega_{Liv}$ and $\omega_{AWIG}$ at the same point. At $t=-0.00985 s$, there is a $2.7\ \sigma$ tension between $\omega_{Liv}$ and $\omega_{GR}$, while at the same point there is a $0.8\ \sigma$ tension between $\omega_{Liv}$ and $\omega_{AWIG}$. At $t=-0.00788 s$, there is a $2.2\ \sigma$ tension with $\omega_{GR}$ but only a $0.5\ \sigma$ tension with $\omega_{AWIG}$. At $t=-0.00591 s$, there is a $1.7\ \sigma$ tension with GR but only a $0.1\ \sigma$ tension with the AWIG prediction. Finally, at $t=-0.00197 s$, there is a $2.1\ \sigma$ tension with GR but only a $0.4\ \sigma$ tension with the AWIG prediction. Therefore, all five data points favour AWIG over GR in this case.
  
\end{subsubsection}

\end{subsection}
\end{section}


\begin{section}{Conclusions}

  In this paper, the model of asymptotically Weyl invariant gravity has undergone several important developments. A potentially exact form for the exponent $n(\mathcal{R})$ is derived, which may complete the definition of the model and shed light on the ubiquitous observation of dynamical dimensional reduction in quantum gravity~\cite{Carlip:2009kf,Carlip:2017eud,Carlip:2019onx}. It is shown how the obtained differential form for $n(\mathcal{R})$ can resolve several outstanding issues with Palatini gravity.


  Perhaps the most important development, however, is the extraction of an experimentally falsifiable prediction capable of distinguishing our model from general relativity. A preliminary analysis of gravitational wave GW150914 yields a maximum tension of $0.9\sigma$ with GR and marginally favours AWIG. A similar study of gravitational wave GW151226 yields a maximum tension of $2.7\sigma$ with GR and more significantly favours AWIG. In fact, 11 out of 11 of the data points analysed favour AWIG over GR, at an average preference level of $1.0\sigma$. This, coupled with the fact that four observations of two different gravitational waves all exhibit similar deviations from general relativity, a general pattern predicted by AWIG, is quite compelling.
  
 However, given the potential importance of this result, a more detailed analysis of the raw gravitational wave data for a large number of binary black hole systems spanning a wide range of total masses is needed. Since other specific models of $f(R)$ gravity may also predict similar modifications of gravitational wave data it is also important to further refine the exact predictions of AWIG.

On the theoretical side, an important next step is to rigorously investigate the renormalizability of AWIG, beyond simple power-counting arguments, in addition to exploring whether or not the theory is truly unitary and capable of globally evading the Ostragadski instability. 
 
\end{section}


\begin{section}{Appendix}
\begin{subsection}{Robustness of prediction}\label{appx1}

  To check how sensitive the AWIG prediction is to the specific ansatz for $n\left(\mathcal{R}_{*}\right)$ we explore two alternative exponents. The first is a general logistic function

\begin{equation}\label{appendix1}
  n\left(\mathcal{R}_{*}\right)=1+\frac{1}{1+ e^{-k(\mathcal{R}_{*}-d)}}.
\end{equation}

For simplicity, we set the transition value to be $d=1$. Note that in this case $\mathcal{R}_{*}$ is no longer restricted to the domain $[0,1]$. We now attempt to set the steepness parameter $k$ such that known experimental limits are satisfied. In Ref.~\cite{Clifton:2005aj} an extended gravitational Lagrangian of the form $R^{1+\delta}$ is considered in the metric formalism. The authors establish a bound of $0\leq \delta \leq 7.2\times 10^{-19}$ based on specific cosmological and solar system constraints~\cite{Clifton:2005aj}. A Palatini action that is linear in the scalar curvature is identical to normal general relativity~\cite{Sotiriou:2008rp,WaldBook}. Therefore, this experimental constraint on $\delta$ in the metric formalism can also be used to constrain the low-curvature value of $n\left(\mathcal{R}_{*}\right)$ to a good approximation. A simple calculation shows that the steepness parameter $k$ must then be bound by $k \geq 43$ if $\delta \leq 7.2\times 10^{-19}$.

Setting $k=43$ and $d=1$, we can compute $T(\mathcal{R}_{*})$ via Eq.~(\ref{b1}) and find the value of $\mathcal{R}_{*}$ for which $T=0$, which we assume corresponds to the high curvature limit of the inspiral process as before. Plugging this value of $\mathcal{R}_{*}$ back into our expression for $f'(\mathcal{R}_{*})$ gives the maximum value by which the metric tensor is rescaled as the two black holes probe the high curvature limit. Figure~(\ref{fig:12ABalog}) shows the value of $f'(\mathcal{R}_{*})$ when $T=0$ for different values of the steepness parameter between $k=1$ and $k=100$, and Figure~(\ref{fig:12ABblog}) shows the value of $f'(\mathcal{R}_{*})$ when $T=0$ for different values of the transition parameter between $d=0.1$ and $d=10$. Thus, the prediction $f'(\mathcal{R}_{*})_{max} \approx 2$ is robust over a wide range of $k$ and $d$ values. For $k \geq 43$ and $d\geq 1$ the factor by which the metric tensor is rescaled is bound by $2.05\geq f'(\mathcal{R}_{*})_{max}\geq 2$.

\begin{figure}[H]
  \centering
    \subfloat[\label{figa}]{\includegraphics[width=8cm]{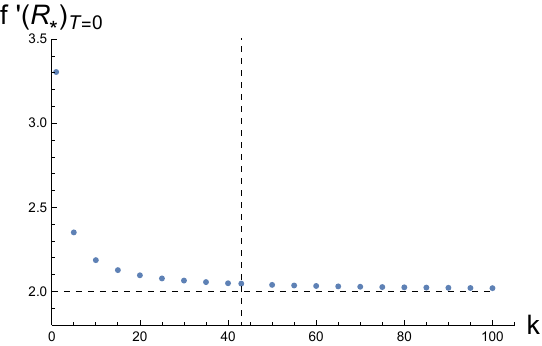} \label{fig:12ABalog}}
  \subfloat[\label{figb}]{\includegraphics[width=8cm]{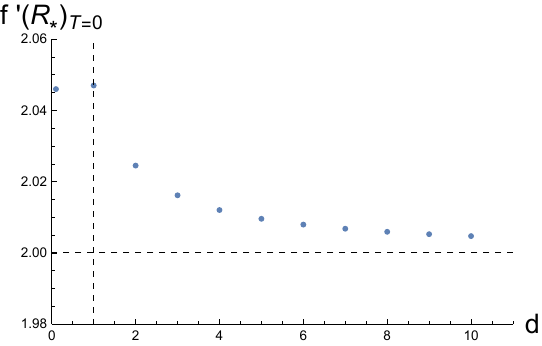} \label{fig:12ABblog}}
  \caption{The value of $f'(\mathcal{R}_{*})$ when $T=0$ as a function of (a) the steepness parameter $k$ for a fixed $d=1$ and (b) the transition parameter $d$ for fixed $k=43$.}
\label{fig:logistic}
\end{figure}

\noindent At lower curvature scales, the minimum value by which the metric tensor is rescaled assuming the ansatz of Eq.~(\ref{appendix1}) with $k=43$ and $d=1$ is $f'(\mathcal{R}_{*})_{min}=0.901$.

Analysing a second alternative ansatz, namely the trigonometric function $n\left(\mathcal{R}_{*}\right)=(1/2)(3-\rm{cos}(\pi \mathcal{R}_{*}))$, gives $f'(\mathcal{R}_{*})_{max}=2$ and $f'(\mathcal{R}_{*})_{min}=0.570$. The deviations from GR predicted by AWIG are therefore fairly robust with respect to the particular function $n\left(\mathcal{R}_{*}\right)$.

\end{subsection}

\begin{subsection}{The effects of filtering and how $\omega(t)$ is calculated}\label{appendix2}

  The frequency $\omega(t)$ has been determined from the $h(t)$ data for GW150914 without including a band-pass filter and using two different methods to check for consistency. The first method $f1$ extrapolates the $h(t)$ data based on half a complete cycle, that is $f1=(2(\Delta t_{1/2}))^{-1}$, where $\Delta t_{1/2}$ is determined by finding the times at which $h(t)$ successively crosses the $h(t)=0$ line. The second method $f2$ extrapolates the $h(t)$ data based on quarter of a cycle, that is $f2=(4(\Delta t_{1/4}))^{-1}$, where $\Delta t_{1/4}$ is determined by finding the time difference between when $h(t)$ is an extremum and when $h(t)=0$. Method $f2$ is the one used in this work. Since the osscilation period varies as a function of time for inspiralling black holes, method $f2$ is expected to be more reliable than $f1$. For comparison, $\omega(t)$ when including the band-pass filter is shown in Fig.~\ref{freqtimecheck} as the curve with error bands. As can be seen in Fig.~\ref{freqtimecheck}, methods $f1$ and $f2$ are fairly consistent with one another and with $\omega(t)$ when including the band-pass filter, within the error budget.

  \begin{figure}[H]
  \centering
  \subfloat{\includegraphics[width=9cm]{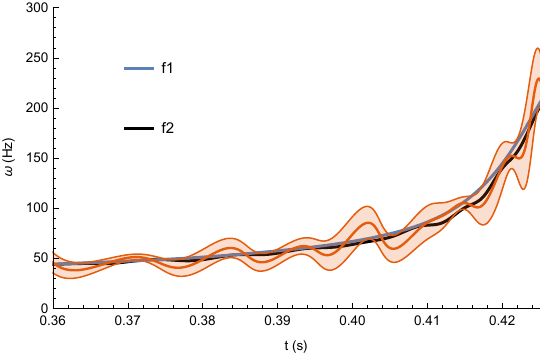}}
  \caption{Central frequency values versus time from numerical relativity simulations of GW150914 without including a band-pass filter~\cite{LIGOScientific:2016aoc}, using two different methods for determining the frequency $f1$ and $f2$. The curve including error bands is $\omega(t)$ when including the band-pass filter.}
  \label{freqtimecheck}
    \end{figure}

 We now repeat the analysis of the Hanford data for GW150914 without applying the filtering to the numerical relativity simulations used in section~\ref{H1} for the two points showing the greatest tension with GR, namely $t=0.405s$ and $t=0.425s$. This is done to gauge the impact of the band-pass filter. However, since the band-pass filter is applied to both the numerical relativity data and the LIGO data the fairest comparison is that presented in section~\ref{H1}. As can be seen in Tab.~\ref{tabAppfin} the results are similar to those presented in section~\ref{H1} but with a reduced statistical significance. Based on this analysis, it seems that adding a band-pass filter only acts to increase the tension with GR. Applying a band-pass filter to the GW151226 data may then actually increase the statistical significance beyond the reported $2.7\sigma$. 

\begin{table}[H]
\begin{center}
\begin{tabular} {|c||c|c|c|c|}
\hline
t(s) & $\omega_{GR}$ & $\omega_{Han}$ & $\omega_{GR}$/$\omega_{Han}$ & $\sigma$ tension with $\omega_{GR}$\\ \hline
\hline
0.40500 & $71.0 \pm 7.8$ & $76.1 \pm 25.0$ & $0.9 \pm 0.3$ & $0.2$ \\ \hline
0.42500 & $200.7 \pm 21.5$ & $175.0 \pm 54.3$ & $1.1 \pm 0.4$  & $0.4$ \\ \hline
\end{tabular}
\end{center}
\caption{A table showing the 2 times at which the frequency predicted by general relativity $\omega_{GR}$ without a band-pass filter is compared with the frequency measured by the Hanford observatory $\omega_{Han}$ for GW150914.}  
\label{tabAppfin}
\end{table}

\end{subsection}
\end{section}


\bibliographystyle{unsrt}
\bibliography{Master}

\end{document}